\documentclass[10pt]{article}
\usepackage{amsmath,amsthm,amsfonts,amssymb,epic,epsfig,graphicx}
\usepackage[dvips]{geometry}
\usepackage{verbatim}
\usepackage{color}

\usepackage{curves}
\usepackage{graphics}
\usepackage{verbatim}

\newcommand{\reals}{\mathbb{R}}

\newcommand{\exponent}{\operatorname{e}}

\theoremstyle{plain}

\newtheorem{prop}{Proposition}

\theoremstyle{remark}
\newtheorem{opm}{Remark}
\begin{document}
\title{Within-host HIV models with periodic antiretroviral therapy}

\author{Patrick De Leenheer\footnote{Department of Mathematics, University of Florida, email: deleenhe@math.ufl.edu}\; \footnote{Supported in part by NSF grant DMS-0614651.}}

\date{}
\maketitle
\begin{abstract}
This paper investigates the effect of drug treatment on the standard within-host HIV model, assuming 
that therapy occurs periodically. It is shown that eradication is possible  
under these periodic regimes, and we quantitatively characterize 
successful drugs or drug combinations, both theoretically and numerically. 
We also consider certain optimization problems, motivated for instance, by the fact that eradication should be achieved at acceptable toxicity levels to the patient. It turns out that these optimization 
problems can be simplified considerably, and this 
makes calculations of the optima a fairly straightforward task. All our results will be illustrated 
by means of numerical examples based on up-to-date knowledge of parameter values in the model.
\end{abstract}

\section{Introduction}
For the past two decades, within-host virus models describing the infection of HIV 
have played an important role in the understanding of this infamous retrovirus, and the ways in which it escapes not only the immune system, but also the various drugs that have been developed 
to suppress viral replication. 
Testing specific hypotheses based on clinical data is difficult since detection techniques of the virus are still far from accurate. This justifies the central role played by mathematical models in this area of  research.

An example of a question that has received considerable attention was whether 
drug treatment fails because of the pre-existence of drug-resistant strains, or by the emergence 
of resistant strains after initiation of drug therapy \cite{bonhoeffer97}.
According to \cite{ribiero00}, the former scenario is more likely. Nevertheless, 
the ability of the virus to mutate quickly, into forms which may be less sensitive to drugs has been,  
and continues to be, the focus of much attention, see recent contributions such as  \cite{ball,MathMed}  that study the behavior of multi-strain  models. 

Other research has gravitated around the fact that the periodic regimen in which 
drugs are taken daily (or more frequently), puts a very high strain on the patient, calling 
for therapies minimizing the treatment burden \cite{kirschner}, 
and also leading to investigations of 
the use of STI's (Structured Treatment Interruptions) \cite{seema,krakovska,ortiz}.

This paper revisits a by now classical model, often referred to as the standard model 
\cite{perelson,nowak}, which is a three-dimensional nonlinear ODE whose state consists  
of the concentrations of healthy CD4+ T cells (the targets of the HIV), infected T cells, and 
viral particles. Upon infection of a healthy T cell, one of the first orders of business is to 
make a copy of the viral RNA, using the enzyme reverse transcriptase. This step, which is error-prone and leads to mutations, can be blocked by a class of drugs called reverse transcriptase (RT) inhibitors. Once the viral copy has been produced, double stranded viral DNA integrates in the cell's nucleus as provirus. 
The usual gene expression now does the rest, and viral proteins are produced according to 
the genetic information encoded in the provirus. These proteins are assembled, mature and ultimately new viruses buds off from the infected cell's surface which go on to infect other T cells. During the  maturation stage the protease enzyme is used to cleave long protein chains, and 
the so-called protease (P) inhibitors, are drugs that target this step. If effective, they 
give rise to defective virus. 

The purpose of this paper is to assess theoretically and quantitatively 
what the impact is of periodic drug treatment on the dynamic behavior of the standard model, 
and in particular to determine what it takes to get rid of 
the infection. Mathematically, we obtain a nonlinear periodic ODE, for which in general 
it is difficult to prove global stability 
and this explains why much research has traditionally resorted to simulations. 
Surprisingly though, 
solutions to the standard model ultimately are bounded by solutions of a monotone system, as pointed out 
by d'Onofrio in \cite{donofrio}, and this allows to conclude global stability for the 
nonlinear periodic model.

We will first consider a simple case, where only RT inhibitors are administered, and where it is assumed 
that the drug is of the bang-bang type, i.e. during a period of the treatment cycle, the drug is 
is either active or inactive. 
The drug is thus characterized by two parameters: its efficiency level when active, and the duration 
of the activity. 
A major role in our analysis is played by the spectral radius of a non-negative matrix 
(the fundamental matrix solution, evaluated over one period, of the linearization at the infection-free 
equilibrium), which 
is shown to possess expected monotonicity properties in terms of the two parameters that 
characterize the drug. Specifically, this spectral radius -which also controls the speed of convergence 
to the infection-free equilibrium-  is lower when the drug is more potent or when it is active longer. 
Equivalently, convergence to the infection-free equilibrium is faster with a more potent drug, or a drug 
whose activity lasts longer. We will see that these results can be generalized to the case of P inhibitors, or 
to a mix of both RT and P inhibitors. This latter scenario 
reflects more closely the standard practice of administering cocktails of drugs to HIV infected patients.

In reality, the efficiency of a drug is not of the bang-bang type. In fact, current research is investigating 
the effect of including pharmacokinetcs into the picture, and has revealed that the efficiency is 
a periodic signal with an initial steep rise right after drug intake, 
followed by a slower decay over a period, see the work of \cite{dixit,rong} for detailed models. 
Therefore, we turn to this more general case, by approximating the 
efficiency by a more general piecewise constant periodic signal. It turns out that the previous results remain valid.

Finally we turn to optimization problems that involve either maximizing the speed of convergence to the 
infection-free equilibrium while making sure that acceptable toxicity levels are not exceeded, or 
by minimizing toxicity levels, while making sure the speed of convergence does not fall below a certain threshold.

All our results will be illustrated by means of numerical examples of 
within-host models whose parameters are chosen in accordance with current prevailing 
knowledge based on clinical data and extensive experimental evidence.
Our results have the potential to suggest which drug, or which combination of drugs, 
are optimal for a given patient. They can also be used to explore the consequences of changing 
the treatment frequency. The investigation of the impact of periodic treatment cycles on multi-strain models, 
or the effect of STI's is the subject of ongoing research.

{\bf Notation}: For matrices $A$ and $B$, $0\leq A$, $0<A$ means that $A$ is a 
(entry-wise) non-negative, positive matrix respectively, and $A\leq B$ means that $0\leq B-A$. A matrix is called 
quasi-positive if all its off-diagonal entries are non-negative. The spectral radius of a matrix $A$ is defined 
as the largest modulus of all eigenvalues of $A$ and will be denoted by $\rho(A)$.

\section{Within-host HIV model with treatment}
We briefly recall the well-known standard model \cite{perelson,nowak}. 
Let
\begin{eqnarray}\label{no-treat}
{\dot T}&=&f(T)-kVT\nonumber \\
{\dot T^*}&=&kVT-\beta T^*\nonumber \\
{\dot V}&=&N\beta T^*-\gamma V,
\end{eqnarray}
where $T$, $T^*$, $V$ denote the concentrations of healthy and infected $T$-cells, and virus particles respectively. 
All parameters are assumed to be positive. 
The parameters $\beta$ and $\gamma$ are the death rates of infected $T$-cells and virus particles respectively. 
The infection is represented by a mass action term $kVT$, and $N$ is the average number of virus particles budding off an infected $T$-cell during its lifetime. The (net) growth rate of the uninfected $T$-cell population is 
given by the smooth function $f(T):\reals_+ \rightarrow \reals$, which is assumed to satisfy the following:
\begin{equation}\label{T0}
\exists \; T_0>0 \; :\; f(T)(T-T_0)<0\textrm{ for } T\neq T_0,\textrm{ and } f'(T_0)<0.
\end{equation}
We have chosen to make the class of allowable $f(T)$'s as large as possible, since the growth 
rate is hard to determine. In addition, most mathematical results apparently remain valid for this large class. Finally, we notice that the two most popular choices for $f(T)$, namely 
$a-bT$ for some positive $a$ and $b$, see \cite{nowak}, and $s+rT(1-T/T_{\max})$ for some positive 
$s,r$ and $T_{\max}$, see \cite{perelson} (here $s$ is a source term modeling $T$ cell production 
in the thymus and $r$ and $T_{\max}$ are the maximal per capita growth rate and carrying capacity 
respectively describing logistic growth of $T$ cells), satisfy the preceding conditions.

Since continuity of $f$ implies that $f(T_0)=0$, it is easy to see that
$$
E_0=(T_0,0,0),
$$
is an equilibrium of $(\ref{no-treat})$, and we will refer to it as the infection-free equilibrium.

A second, positive equilibrium (corresponding to an infection) may 
exist if the following quantities are positive:
\begin{equation}\label{eq}
{\bar T}=\frac{\gamma}{kN},\;\; {\bar T^*}=\frac{f({\bar T})}{\beta},\;\; {\bar V}=\frac{f({\bar T})}{k{\bar T}}.
\end{equation}
Note that this is the case iff $f(\frac{\gamma}{kN})>0$, or equivalently by $(\ref{T0})$ 
that ${\bar T}=\frac{\gamma}{kN}<T_0$. In terms of the basic reproduction number
$$
R_0:=\frac{kN}{\gamma}T_0,
$$
existence of a positive equilibrium is therefore equivalent with 
\begin{equation}\label{standing}
1<R_0,
\end{equation}
which will be a standing assumption throughout the rest of this paper. Indeed, if we would assume that 
$R_0<1$, it is known from \cite{SIAP03} that the infection-free equilibrium $E_0$ is globally asymptotically stable (GAS), 
and hence in this case the infection would always be cleared without treatment. 

We denote the positive equilibrium that corresponds to an infection by 
$E=({\bar T},{\bar T^*},{\bar V})$. 
Linearization at $E_0$ shows that it is unstable, and 
conditions on $f(T)$ are known that guarantee that $E$ is GAS
(excluding of course initial conditions corresponding to a healthy, uninfected individual; 
these coincide with the $T$-axis, which is the stable manifold of $E_0$). 
However, it is also possible that the model exhibits sustained oscillatory solutions 
which can be asymptotically stable. Regardless of the dynamical complexity of the solutions of the model, in general, 
if left untreated, the infection will persist within a patient. 
All these results follow from \cite{SIAP03}. 

Obviously, the purpose of treatment is to clear the infection, hopefully by making 
$E_0$ GAS by suitable modifications of model $(\ref{no-treat})$ which reflect the effect of 
drugs. For the moment, we will only consider the effect of RT inhibitors, but P inhibitors will be included later. 
Using monotherapy based on RT inhibitors, model $(\ref{no-treat})$ is modified to:

\begin{eqnarray}\label{treat}
{\dot T}&=&f(T)-k(1-\epsilon(t))VT\nonumber \\
{\dot T^*}&=&k(1-\epsilon(t))VT-\beta T^*\nonumber \\
{\dot V}&=&N\beta T^*-\gamma V,
\end{eqnarray}
where $\epsilon(t)\in [0,1]$ is the (time-varying) drug efficiency of the RT 
inhibitors. The drug is not effective when $\epsilon(t)=0$ and $100 \%$ effective when $\epsilon(t)=1$. 
Notice that $E_0$ is still an equilibrium of the modified model $(\ref{treat})$, regardless of the drug efficiency. 

Assuming that the efficiency is constant over time, we set $\epsilon(t)=e \in (0,1]$. 
Then, to clear the infection, it suffices to choose $e$ such that the modified basic reproduction number $R_0(\epsilon)$ 
is less than $1$, where
$$
R_0(\epsilon):=\frac{k(1-e)N}{\gamma}T_0.
$$
Indeed, the results mentioned previously are applicable to this modified model, and they imply that if 
$R_0(\epsilon)<1$, then $E_0$ is GAS for $(\ref{treat})$. Equivalently, if the efficiency $e$ satisfies 
$$
e>1-\frac{1}{R_0},
$$
then treatment will be successful in this case. If the drug would be effective $100 \%$ so that $e=1$, then 
treatment would always be successful. Current RT inhibitors clearly do not fit this profile. Moreover, 
in practice, the drug efficiency is not constant through time, and the main purpose of this 
paper is to investigate the quantitative consequences of this fact.

\section{Periodic drug efficiency}
We now make the assumption that $\epsilon(t)$ is periodic:
$$
\epsilon(t)=\epsilon(t+\tau),\textrm{ for all } t,
$$
for some period $\tau>0$. This is closer to reality where 
patients ideally adhere to a strict periodic treatment schedule, taking 
medication daily ($\tau=1$ day) or twice a day ($\tau=0.5$ day) for instance.

The shape of $\epsilon(t)$ 
over one period is determined by the (here unmodeled) pharmacokinetics, although 
coupling of the standard model with detailed pharmacokinetics models  
has been the subject of recent research, see for instance the work of \cite{dixit,rong},  
\begin{figure}
\centerline{
\includegraphics[width=11cm]{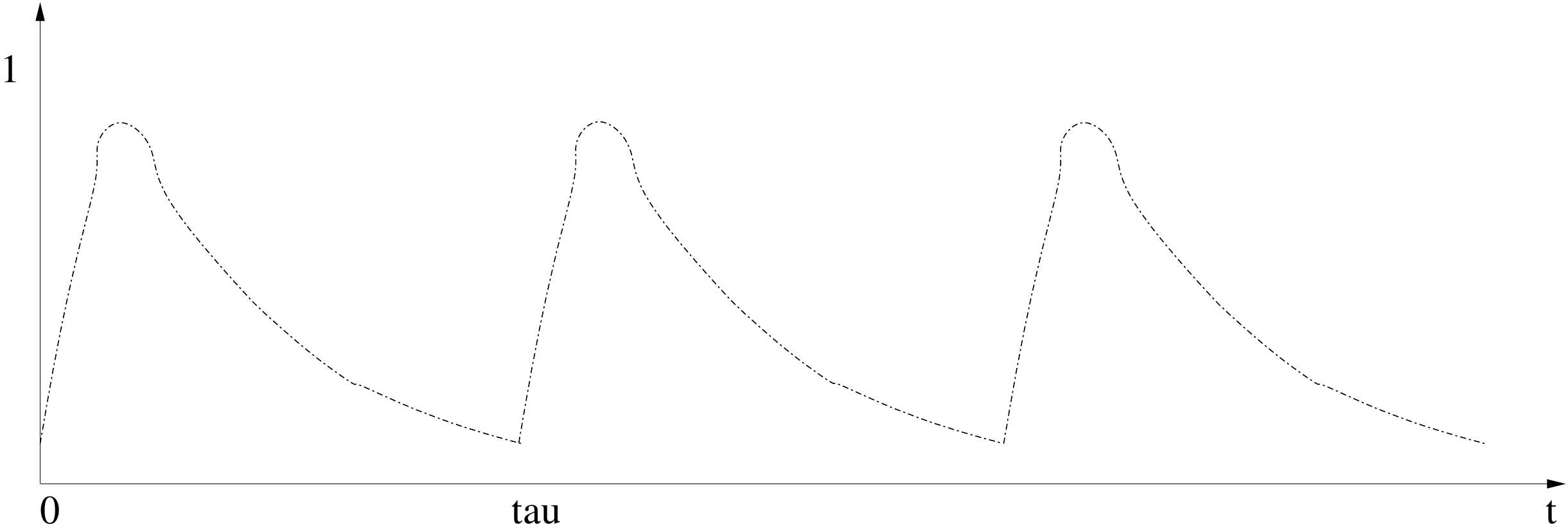}}
\caption{Periodic drug efficiency $\epsilon (t)$.}\label{ef0}
\end{figure}
where it was shown that at least qualitatively, the graph of the periodic function $\epsilon(t)$ is 
roughly like the one depicted in Figure $\ref{ef0}$. It is characterized by 
a quick rise of the efficiency to a peak value right after drug intake, followed by a slower decay.  
This is significantly different from the case where the efficiency is constant, the situation we 
described in the previous section. In pharmacokinetics, the efficiency $\epsilon (t)$ is traditionally defined 
as 
$$
\epsilon(t)=\frac{y(t)}{K+y(t)},
$$
for some positive constant $K$. Here, $y(t)$ is an output of a linear compartmental system 
$$
{\dot z}=Mz(t)+u(t),\;\; y(t)=z_n(t),
$$
where $M$ is a stable compartmental matrix (i.e. $M$ is an $n\times n$ quasi-positive matrix whose eigenvalues 
are in the open left half plane), and the state components describe the concentrations of 
the drug in various compartments of the model (gut, blood, etc). Typically, $y(t)$ is the 
concentration of an activated form of the drug inside the target cells. The input $u(t)$ describes the 
(ideally periodic) drug intake signal and it is often modeled as a pulse. It is not difficult to show that 
for periodic $u(t)$, the output $y(t)$ will converge to a periodic signal with the same period, and this justifies 
to assume that $\epsilon(t)$ is also periodic with the same period. 

Assuming a periodic efficiency $\epsilon(t)$, let us start by linearizing 
system $(\ref{treat})$ at the equilibrium $E_0$:
\begin{equation}\label{linearization}
{\dot x}=B(t)x,
\end{equation}
where
$$
B(t)=\begin{pmatrix} 
f'(T_0)&0&-k(1-\epsilon (t))T_0\\
0&-\beta&k(1-\epsilon (t))T_0\\0&N\beta & -\gamma
\end{pmatrix}.
$$
It is well-known that the stability properties of the origin of $(\ref{linearization})$ 
(and generically the local stability properties of the equilibrium $E_0$ for system $(\ref{treat})$) 
are determined by the Floquet multipliers of $(\ref{linearization})$. The block-triangular structure of 
$B(t)$ implies that these are
$$
\exponent^{f'(T_0)\tau}\textrm{ and } \lambda_2,\lambda_3,
$$
where $\lambda_2$ and $\lambda_3$ are the Floquet multipliers of the planar 
$\tau$-periodic system:
\begin{equation}\label{reduced}
\begin{pmatrix}{\dot x_2}\\{\dot x_3}\end{pmatrix}
=\begin{pmatrix} -\beta & k(1-\epsilon (t))T_0\\N\beta&-\gamma \end{pmatrix}
\begin{pmatrix}x_2\\x_3 \end{pmatrix}
\end{equation}
In particular, since $f'(T_0)<0$ by $(\ref{T0})$, it follows that the three 
Floquet multipliers of system $(\ref{linearization})$ are contained in 
the interior of the unit disk of the complex plane -which in turn 
implies that $E_0$ is locally asymptotically stable  
for system $(\ref{treat})$- if $|\lambda_2|,|\lambda_3|<1$. 
In fact, by a beautiful argument due to d'Onofrio in \cite{donofrio}, it turns out that the same conditions 
imply the much stronger result of {\it global} asymptotic stability of $E_0$ for system $(\ref{treat})$.
\begin{prop}\cite{donofrio} \label{dono}
Let the Floquet multipliers of system $(\ref{linearization})$ be contained in the interior of the 
open unit disk of the complex plane. Then $E_0$ is GAS for system $(\ref{treat})$, hence the 
infection is cleared.
\end{prop}
This result shows how relevant and important it is to determine the Floquet multipliers of system 
$(\ref{linearization})$. Unfortunately, for general functions $\epsilon (t)$, this is a notoriously 
difficult task. Therefore, we will consider the simpler case where 
$\epsilon(t)$ is piecewise constant, bearing in mind that piecewise constant functions are often 
good approximations to continuous functions. We will start with an even simpler case where 
$\epsilon(t)$ is of the bang-bang type.

\begin{figure}
\centerline{
\includegraphics[width=7cm]{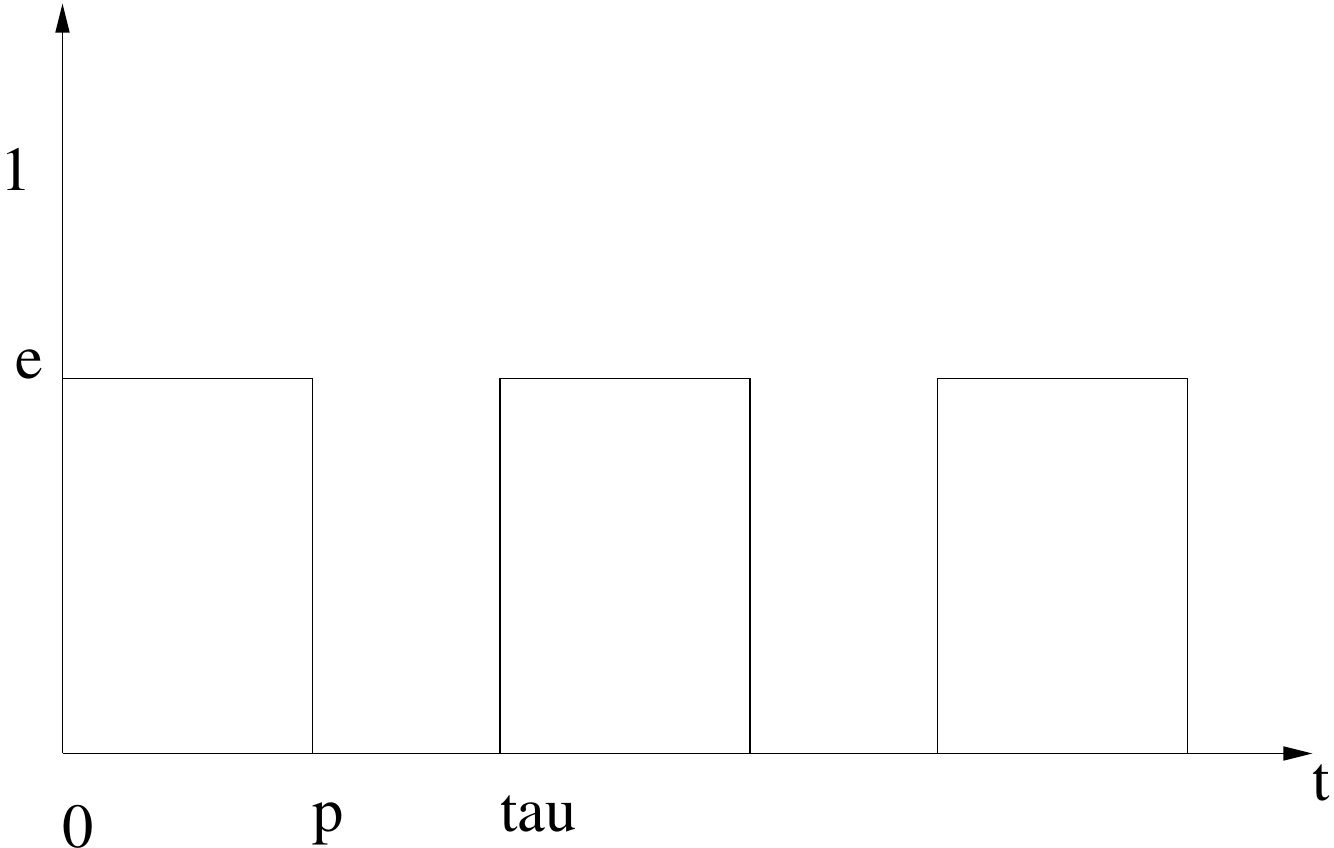}}
\caption{Periodic drug efficiency $\epsilon (t)$ of the bang-bang type.}\label{ef}
\end{figure}

\subsection{Periodic drug efficiency of the bang-bang type}
We make the following simplifying assumption regarding 
the shape of the graph of the $\tau$-periodic function $\epsilon (t)$, which is illustrated in Figure 
$\ref{ef}$: 
\begin{equation}\label{efficiency}
\epsilon (t)=\begin{cases}e,\;\; t\in [0,p]\\
0,\;\; t\in (p,\tau)\end{cases},
\end{equation}
where $p\in (0,\tau)$ is the time duration during which the drug is 
supposed to be active with efficiency $e\in [0,1]$.  
During the remaining part of the treatment period the drug is assumed to be totally 
inefficient. Clearly, this is a very crude way of approximating the more realistic shape of 
$\epsilon(t)$ depicted in Figure $\ref{ef0}$, but some key properties are 
to be learned from this case, and they carry over to more general cases that describes reality better, as 
we will discover later.

There are two possible parameters which can be varied in 
$(\ref{efficiency})$, namely $e$ and $p$, and the purpose of the rest of 
this subsection is to investigate their effect on the Floquet multipliers 
of system $(\ref{reduced})$ with $(\ref{efficiency})$. 
These Floquet multipliers are the eigenvalues of the following matrix
\begin{equation}\label{def-fi}
\Phi(e,p)=\exponent^{(\tau-p)A(0)}\exponent^{pA(e)}
\end{equation}
where 
\begin{equation}\label{A-mat}
A(e):=\begin{pmatrix}-\beta &  k(1-e)T_0\\
N\beta & -\gamma\end{pmatrix}.
\end{equation}
Since both $A(e)$ and $A(0)$ are quasi-positive matrices 
their matrix exponentials are non-negative matrices 
\footnote{Proof: Let $A$ be quasi-positive. Then $B=A+\alpha I$ is a non-negative matrix 
for all sufficiently large values of $\alpha$, implying that 
$\exponent^{tB}$ is a non-negative matrix for all $t\geq 0$. But since $\exponent^{tA}=
\exponent^{-\alpha t}\exponent^{tB}$, the same conclusion holds for $\exponent^{tA}$.} 
Thus, $\Phi(e,p)$ is a non-negative matrix and by the Perron-Frobenius Theorem \cite{berman-plemmons} 
its spectral radius $\rho \left(\Phi(e,p)\right)$ is an eigenvalue of $\Phi(e,p)$. Thus, the Floquet multipliers   
of system $(\ref{reduced})$ with $(\ref{efficiency})$ are contained in the interior 
of the unit disk of the complex plane if and only if $\rho \left(\Phi(e,p) \right)<1$. 
This guarantees that the infection is cleared (globally) by Proposition $\ref{dono}$.

\begin{figure}
\centerline{
\includegraphics[width=7cm]{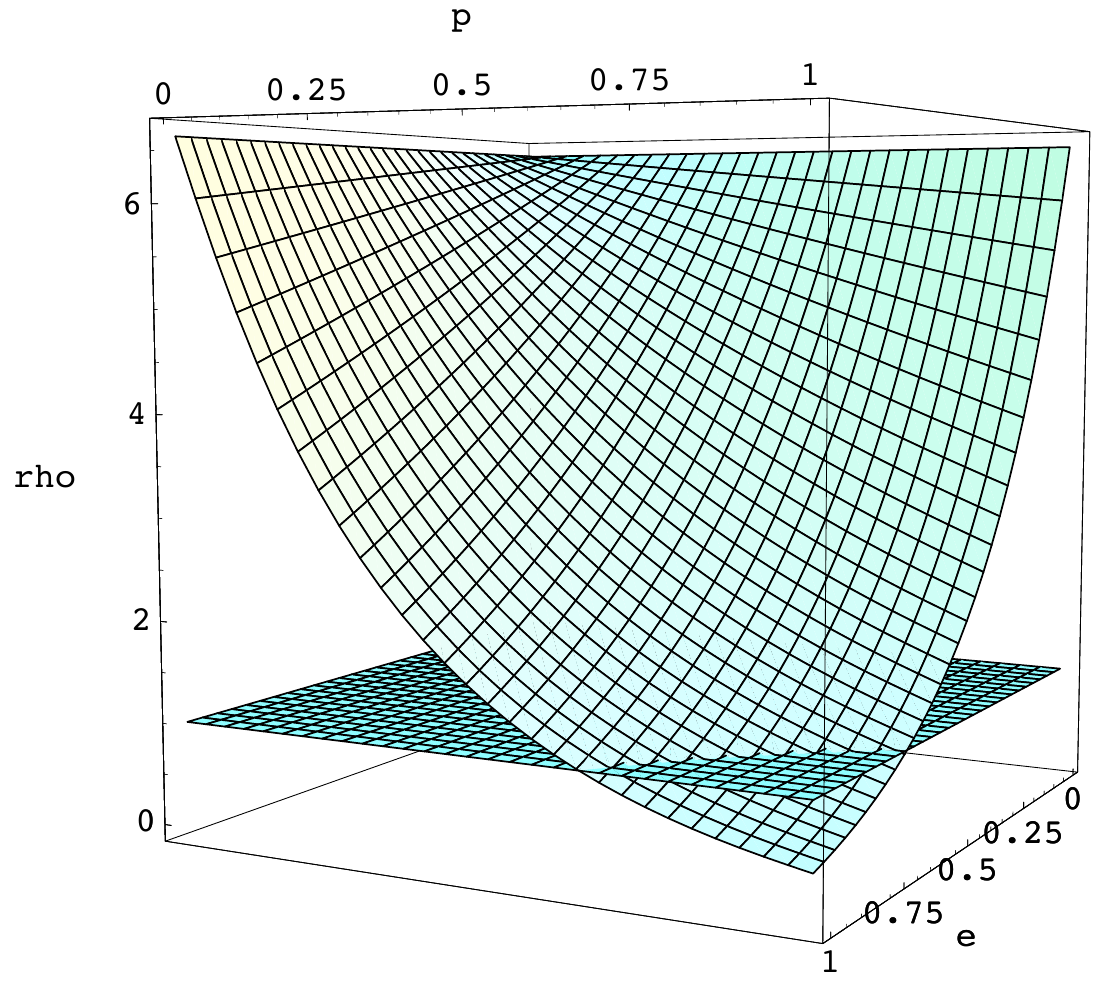}}
\caption{Spectral radius of $\rho\left(\Phi(e,p)\right)$ as a function of efficiency $e$ and treatment duration $p$. The horizontal surface corresponds to $\rho=1$.}\label{3d}
\end{figure}

\begin{figure}
\centerline{
\includegraphics[width=7cm]{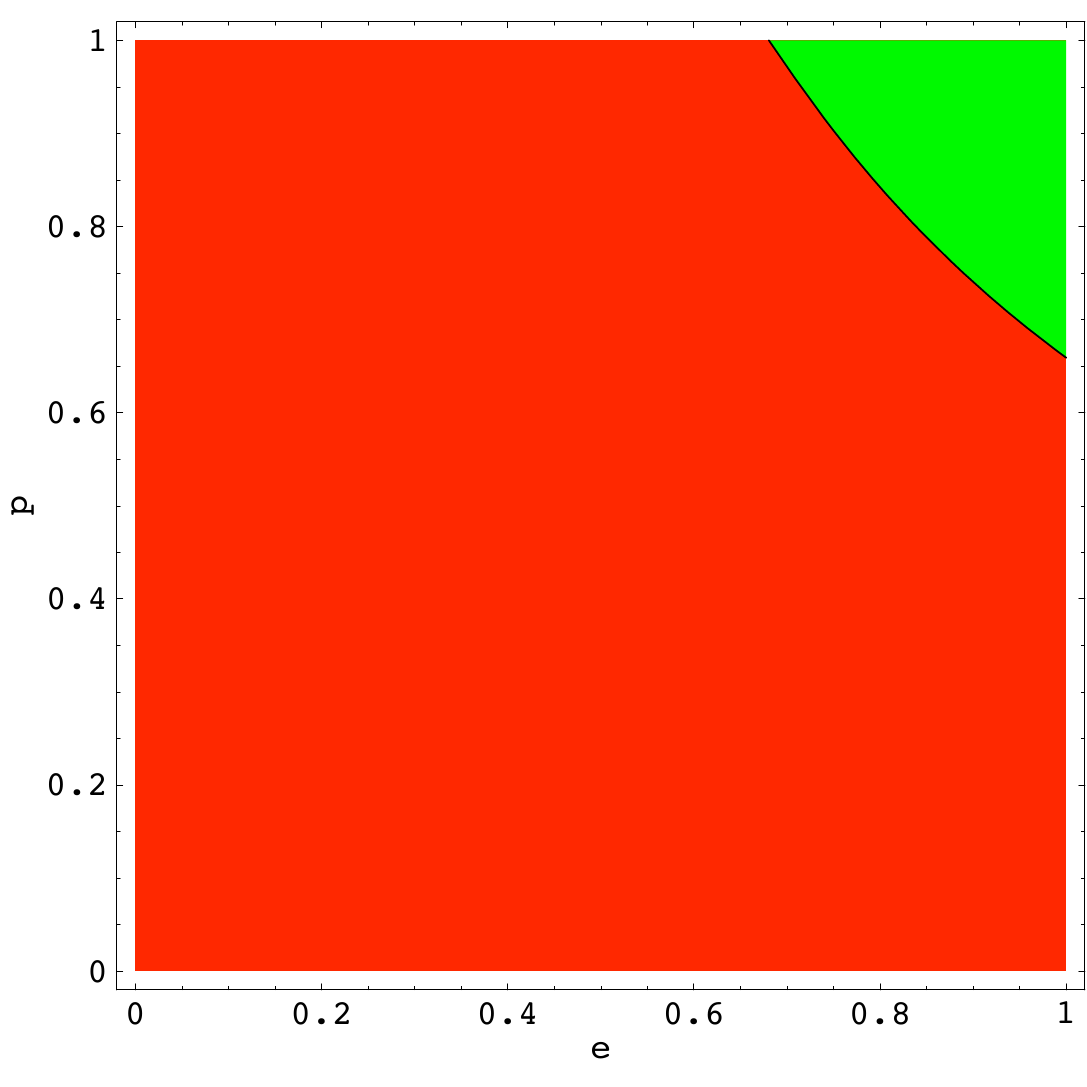}}
\caption{Contour plot of the spectral radius $\rho\left(\Phi(e,p)\right)$ 
as a function of efficiency $e$ and treatment duration $p$: $\rho\left(\Phi(e,p) \right)>1$ in red region and $<1$ in green region.}\label{cont}
\end{figure}

The following proposition -whose proof is deferred to the Appendix- 
reveals that $\rho \left(\Phi(e,p)\right)$ has 
the expected monotonicity properties: it decreases with $e$ 
(more efficient treatment) and with $p$ (drug is effective longer). 
\begin{prop}\label{mono}
Let $e,e'\in [0,1]$ and $p,p'\in [0,\tau]$. Then the map 
$(e,p)\rightarrow \rho\left(\Phi(e,p)\right)$ is continuous, 
\begin{equation}\label{mon-eps}
e<e',\;\; p\neq 0\Rightarrow \rho \left(\Phi(e',p)\right)<
\rho \left(\Phi(e,p)\right),
\end{equation}
and 
\begin{equation}\label{mon-p}
p<p',\;\;e\neq 0\Rightarrow \rho \left(\Phi(e,p')\right)<
\rho \left(\Phi(e,p)\right).
\end{equation}
Moreover,
\begin{equation}\label{boundary}
\rho(\Phi(e,0))=\rho(\Phi(0,\tau))=\rho(\exponent^{\tau A(0)})>1\textrm{ for all } 
e \in [0,1]\textrm{ and all } p\in [0,\tau]\textrm{ (no treatment)}.
\end{equation}
and 
\begin{equation}\label{boundary2}
\rho\left(\Phi(1,\tau) \right)=\rho \left(\exponent^{\tau A(1)}\right)=
\max \{\exponent^{-\beta \tau},\exponent^{-\gamma \tau} \}<1\textrm{ (constant, $100 \%$ effective treatment).}
\end{equation}
\end{prop}
Since $\rho \left( \Phi (e,p)\right)$ (provided it is less than $1$) is 
a measure of how fast solutions of $(\ref{treat})$ with $(\ref{efficiency})$ approach $E_0$ (at least 
locally near $E_0$), this result may be interpreted as follows: 
\begin{center}
{\it Let the treatment be periodic, of the bang-bang type, and capable of clearing the infection.
If it is more efficient, or lasts longer, then the infection is cleared more quickly.} 
\end{center}
We illustrate Proposition $\ref{mono}$ in Figures $\ref{3d}$ and $\ref{cont}$. The parameters used are taken from \cite{rong}, and they are as follows: $f(T)=a-bT$ with $a=10^4 \textrm { ml}^{-1}\textrm{ day}^{-1}$ and 
$b=0.01\textrm{ day}^{-1}$ (which implies that $T_0=10^6\textrm{ ml}^{-1}$), 
$k=2.4\times 10^{-8}\textrm{ ml day}^{-1}$, $\beta=1\textrm{ day}^{-1}$, $N=3000$, 
$\gamma =23\textrm{ day}^{-1}$. The period of the treatment $\tau$ is $1$ day.

\begin{opm}
This result can be modified to the situation in which P inhibitors  are 
used for treatment instead of RT inhibitors. Model  $(\ref{treat})$ is then replaced by \begin{eqnarray}\label{treat2}
{\dot T}&=&f(T)-kVT\nonumber \\
{\dot T^*}&=&kVT-\beta T^*\nonumber \\
{\dot V}&=&N(1-\epsilon(t))\beta T^*-\gamma V,
\end{eqnarray}
and matrix $A(e)$ in $(\ref{A-mat})$ by
\begin{equation}\label{A-mat2}
A(e)=\begin{pmatrix}-\beta &  kT_0\\
N(1-e)\beta & -\gamma\end{pmatrix}.
\end{equation}
With this notation and still using $(\ref{def-fi})$, Proposition $\ref{mono}$ remains valid.
\end{opm}
\begin{opm}
Similar results can be stated to describe the situation in which combination therapy 
is used. This is the more commonly found therapy method where patients take a cocktail of both 
RT and P inhibitors.
Model $(\ref{treat})$ should then be replaced by 
\begin{eqnarray}\label{treat3}
{\dot T}&=&f(T)-k(1-\epsilon_{RT}(t))VT\nonumber \\
{\dot T^*}&=&k(1-\epsilon_{RT}(t))VT-\beta T^*\nonumber \\
{\dot V}&=&N(1-\epsilon_{P}(t))\beta T^*-\gamma V,
\end{eqnarray}
where 
\begin{equation}\label{efficiency2}
\epsilon_{RT} (t)=\begin{cases}e_{RT},\;\; t\in [0,p_{RT}]\\
0,\;\; t\in (p_{RT},\tau)\end{cases},\;\; 
\epsilon_{P} (t)=\begin{cases}e_{P},\;\; t\in [0,p_{P}]\\
0,\;\; t\in (p_{P},\tau)\end{cases}
\end{equation}
denote the piecewise constant efficiencies of the RT and P inhibitors respectively. Finally, matrix $A(e)$ in $(\ref{A-mat})$ is replaced by
\begin{equation}\label{A-mat3}
A(e_{RT},e_{P})=\begin{pmatrix}-\beta &  k(1-e_{RT})T_0\\
N(1-e_{P})\beta & -\gamma\end{pmatrix}.
\end{equation}
With these notations and assuming without loss of generality that $p_{RT}<p_{P}$ (if not, simply swap subscripts 
RT and P in the expression below), the spectral radius of the following matrix
$$
\Phi(e_{RT},e_{P},p_{RT},p_{P})=\exponent^{(\tau-p_{P})A(0,0)}\exponent^{(p_{P}-p_{RT})A(0,e_{P})}\exponent^{p_{RT}A(e_{RT},e_{P})}
$$
is the key quantity. As expected, the spectral radius 
is increasing in each of its arguments $e_{RT},e_{P},p_{RT},p_{P}$. We omit the proofs of these results 
as they are straightforward modifications of the proof of Proposition $\ref{mono}$.
\end{opm}

\subsection{General piecewise constant periodic drug efficiencies}
\begin{figure}
\centerline{
\includegraphics[width=7cm]{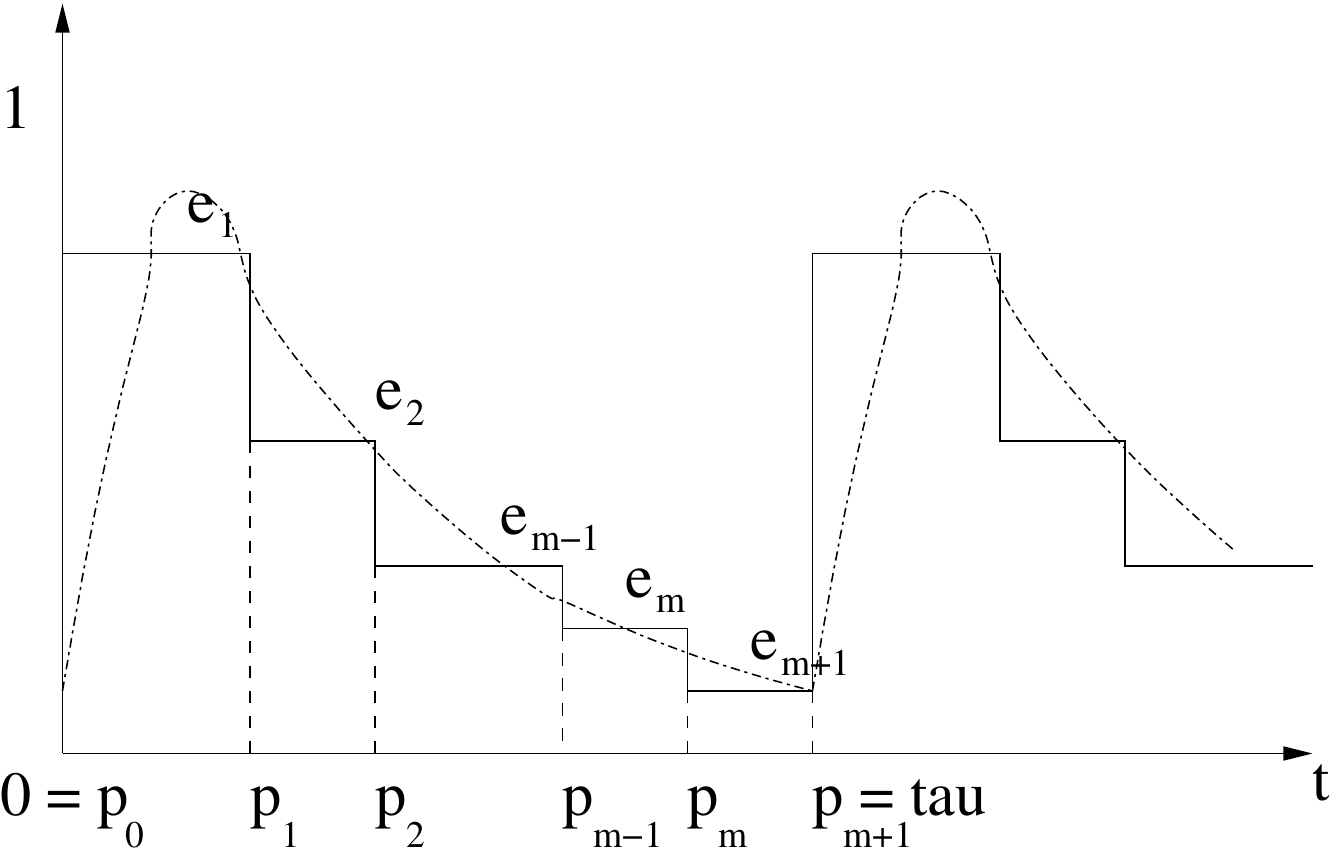}}
\caption{$\tau$-periodic drug efficiency $\epsilon (t)$ (dashed-dotted line) and a piecewise constant approximation.}\label{ef2}
\end{figure}

As mentioned earlier, in practice, the graph of the drug efficiency is not as shown in Figure $\ref{ef}$, but rather as the dashed-dotted line in Figure $\ref{ef2}$, which can be approximated by a piecewise constant and $\tau$-periodic efficiency with several constant  
drug level efficiencies $e_1>e_2>\dots>e_m>e_{m+1}$ during the respective 
intervals $[p_0,p_1),[p_1,p_2),\dots,[p_{m-1},p_m),[p_m-p_{m+1})$, 
where $p_0:=0$ and $p_{m+1}:=\tau$ for some $m\geq 1$.

Define $(E,P):=(e_1,\dots,e_{m+1},p_1,\dots,p_m)$ and let
\begin{equation}\label{general-fi}
\Phi \left(E,P\right)=\exponent^{(p_{m+1}-p_m)A(e_{m+1})}\dots \exponent^{(p_1-p_0)A(e_1)}.
\end{equation}

Similarly to Proposition $\ref{mono}$, we find that the spectral radius of $\Phi(E,P)$ is increasing in 
each of its arguments.

\begin{prop}\label{extension}
Let $m\geq 1$, $0=p_0<p_1<p_2<\dots <p_m<p_{m+1}=\tau$ and $1=e_0\geq e_1>e_2>
\dots>e_m>e_{m+1}\geq e_{m+2}=0$. Then the map $(E,P)\rightarrow \rho\left(\Phi\left(E,P\right)\right)$ is continuous. In addition,
\begin{equation}\label{ext-eps}
i\in\{1,\dots,m+1\} \textrm{ and }e_{i+1}<e_i<e_i'<e_{i-1}\footnote{If $i=1$, then replace $<$ by $\leq$ in the right most inequality. If $i=m+1$, replace $<$ by $\leq$ in the left most inequality.}
\Rightarrow \rho \left(\Phi \left({\tilde E}_i,P\right) \right)<
\rho \left(\Phi\left(E,P\right)\right),
\end{equation}
and 
\begin{equation}\label{ext-p}
j\in\{1,\dots,m\} \textrm{ and }p_{j-1}<p_j<p'_j<p_{j+1}\Rightarrow \rho \left(\Phi\left(E,{\tilde P}_j\right)\right)<
\rho \left(\Phi\left(E,P\right)\right).
\end{equation}
where ${\tilde E}_i=(e_1,e_2,\dots,e_{i-1},e_i',e_{i+1},\dots,e_{m+1})$ and 
${\tilde P}_j=(p_1,p_2,\dots,p_{j-1},p_j',p_{j+1},\dots,p_m)$.
\end{prop}
The proof is deferred to the Appendix.

\section{Optimization problems}
In this section we return to the case of periodic efficiencies of the bang-bang type. What follows can easily 
be generalized to the case of more general, piecewise constant periodic efficiencies. As mentioned earlier, 
the purpose of treatment is to eradicate the infection by making $E_0$ GAS for $(\ref{treat})$ with $(\ref{efficiency})$. 
In practice however, one would like to achieve this while the burden to the patient is as low as possible. 
Obviously, there are various ways to measure this burden. Let us list a couple of particular problems, 
assuming a $\tau$-periodic treatment schedule:
\begin{enumerate}
\item
Minimize $\rho \left(\Phi(e,p) \right)$ subject to $(e,p)\in [0,1]\times [0,\tau]$ and 
$\int_0^\tau \epsilon(t)dt=ep\leq c$, for some fixed $c\in (0,\tau)$.
\item
Minimize $\int_0^\tau \epsilon(t)dt=ep$, subject to $(e,p)\in [0,1]\times [0,\tau]$ and 
$\rho \left(\Phi(e,p) \right)\leq \delta$, for some fixed $\delta \in (\rho \left( \Phi(1,\tau)\right),1)$.
\end{enumerate}
In the first problem the spectral radius of $\Phi(e,p)$ is minimized. As we mentioned before 
this spectral radius controls the rate of convergence to $E_0$ (provided it is less than $1$): the smaller 
the spectral radius, the faster solutions converge.  
In addition to minimizing the spectral radius, the burden on the patient should not exceed 
a specified upper bound $c$. Here, the burden to the patient is measured as the 
area under the graph of the efficiency $\epsilon(t)$ over one period. The second problem 
on the other hand, concerns minimization of the patient's burden, subject to the condition that the 
spectral radius is less than a given bound $\delta$ (assumed to be less than $1$ so that convergence 
to $E_0$ is guaranteed). 

Both problems fit in the larger classes of problems which we describe next. 
Let the maps $F,G:[0,1]\times[0,\tau]\rightarrow \reals$ be continuously differentiable 
with the following properties:
$$
F(0,0)=G(0,0)=0,\textrm{ and }\nabla F,\; \nabla G \geq 0, \textrm{ but } \neq 0 
\textrm{ on } [0,1]\times[0,\tau]\setminus \{(0,0)\}.
$$
Now consider the more general optimization problems:
\begin{center}
Class I. Minimize $\rho \left(\Phi(e,p) \right)$ subject to $(e,p)\in [0,1]\times [0,\tau]$ and 
$F(e,p)\leq c$, for some fixed $c>0$ 
satisfying $\{(e,p)|F(e,p)=c\}\cap [0,1]\times[0,\tau]\neq \emptyset$.
\end{center}
\begin{center}
Class II. Minimize $G(e,p)$, subject to $(e,p)\in [0,1]\times [0,\tau]$ and 
$\rho \left(\Phi(e,p) \right)\leq \delta$, for some fixed $\delta \in (\rho \left(\Phi(1,\tau) \right),1)$.
\end{center}
The first two problems fit in this class for the choices $F(e,p)=G(e,p)=ep$. But it is clear that other 
choices could be of interest as well, for instance $F(e,p)=ae^{q_1}+bp^{q_2}$, 
for some fixed $q_1,q_2\geq 1$ and $a,b>0$, or positive linear combinations of several of these functions.

It turns out that both classes of optimization problems can be simplified thanks to Proposition $\ref{mono}$: We will 
see shortly that the optimum appears on the boundary of the constraint set in both cases, 
which translates into saying that the optimum occurs only if the patient's burden is the maximally allowed 
one (for problems in the first class), or that the spectral radius takes the largest allowed 
value (for problems in the second class) implying that convergence to $E_0$ will be as slow as allowed.

To be more precise, we claim that Class I and II optimization problems are equivalent to Class III and IV problems respectively which 
are defined as follows:
\begin{center}
Class III. Minimize $\rho \left(\Phi(e,p) \right)$ subject to $(e,p)\in [0,1]\times [0,\tau]$ and 
$F(e,p)= c$, for some fixed $c>0$ satisfying that 
$\{(e,p)|F(e,p)=c\}\cap [0,1]\times[0,\tau]\neq \emptyset$.
\end{center}
\begin{center}
Class IV. Minimize $G(e,p)$, subject to $(e,p)\in [0,1]\times [0,\tau]$ and 
$\rho \left(\Phi(e,p) \right)= \delta$, for some fixed $\delta \in (\rho \left(\Phi(1,\tau) \right),1)$.
\end{center}
Notice that the difference between Class I and III, and Class II and IV is in the constraint only 
(by replacing the inequality by an equality). 
In other words, the optimum of Class I and II problems occurs on the boundary of the constraint set. 
We show this equivalence for Class I and III problems. The argument to show equivalence of 
Class II and IV problems is very similar and omitted. 
Suppose that $(e^*,p^*)$ is such that $\rho(\Phi(e^*,p^*))$ is minimal, while 
$F(e^*,p^*)<c$. Notice that $(e^*,p^*)\neq(0,0)$ since $c>0$ 
and $F$ takes small positive values near $(0,0)$ 
in the rectangular region $R:=[0,1]\times [0,\tau]$ and $\rho$ is strictly lower in those points. 
Also $(e^*,p^*)\neq (1,\tau)$ since otherwise the level set 
$\{(e,p)|F(e,p)=c\}$ does not intersect $R$, contrary to our assumption.

If $(e^*,p^*)$ is in the interior of $R$, then the point 
$(e',p'):=(e^*,p^*)+s\nabla F(e^*,p^*)$ is still in the interior of $R$ with $F(e',p')<c$ 
for small enough positive $s$, yet $\rho \left( \Phi (e',p')\right)<\rho \left(\Phi(e^*,p^*) \right)$ by 
Proposition $\ref{mono}$, contradicting minimality. The same argument applies if $(e^*,p^*)=(0,p^*)$ for 
some $p^*\in (0,\tau)$ or if $(e^*,p^*)=(e^*,0)$ for some $e^*\in (0,1)$, since a perturbation of such a point in the direction 
of $\nabla F(e^*,p^*)$, results in a point which is still in $R$. If $(e^*,p^*)=(1,p^*)$ for some 
$p^*\in (0,\tau)$ or if $(e^*,p^*)=(e^*,\tau)$ for some $e^*\in (0,1)$, then a perturbation in the 
direction of $\nabla F(e^*,p^*)$ could potentially result in a point outside $R$. To prevent this we perturb as follows for the case where $(e^*,p^*)=(1,p^*)$ 
(the argument when $(e^*,p^*)=(e^*,\tau)$ is similar and omitted): 
Let $(e',p')=(1,p^*)+(0,s)$. Then for sufficiently small and positive $s$, $(e',p')$ is still on the 
boundary of $R$ and $F(e',p')<c$, yet $\rho \left(\Phi(e',p')\right)< \rho \left(\Phi(e,p) \right)$ 
by Proposition $\ref{mono}$, a contradiction to minimality.

\section{Numerical examples}
Here we provide some examples of the optimization problems we just discussed. 
The model parameters used throughout this section are the ones chosen 
in subsection $3.1$.

\begin{figure}
\centerline{
\includegraphics[width=7cm]{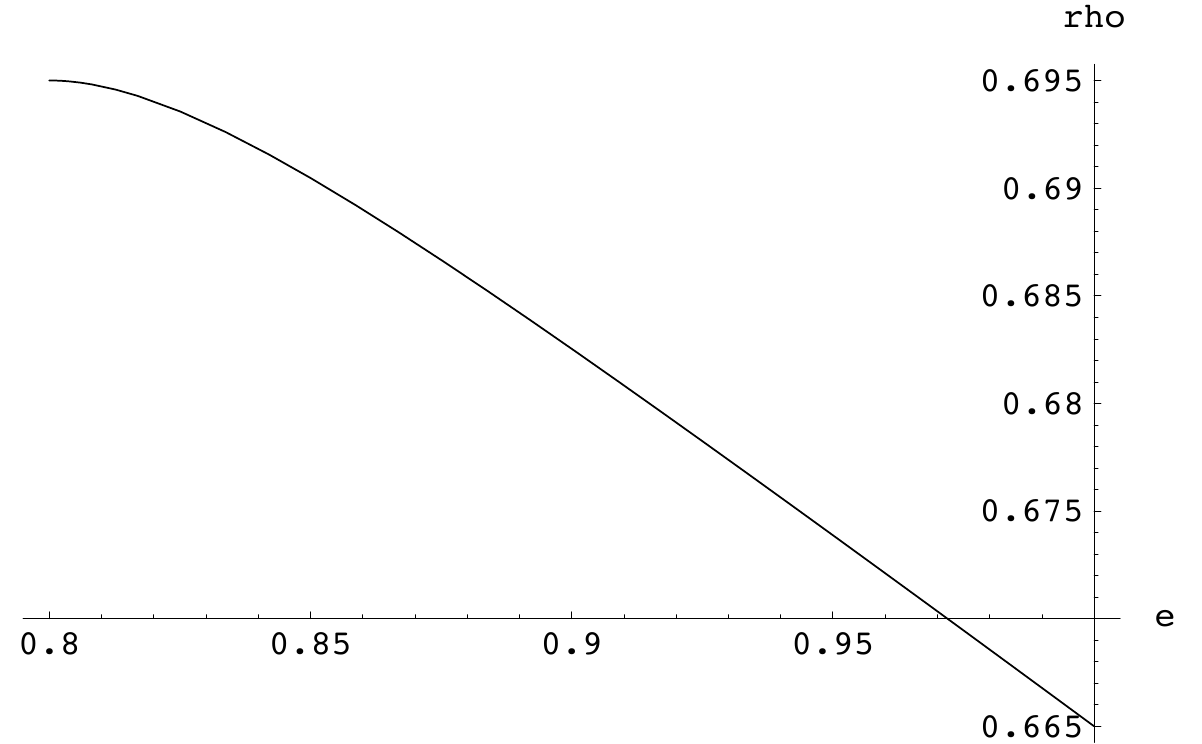}}
\caption{Graph of $\rho \left( \Phi(e,p)\right)$ for points $(e,p)$ with $ep=0.8$ for $e\in[0.8,1]$. Minimum 
for $\rho$ is $0.665$ and it is achieved for $(e,p)=(1,0.8)$.}\label{num1}
\end{figure}

Let us first minimize $\rho \left( \Phi(e,p)\right)$, see Figures $\ref{3d}$ and $\ref{cont}$. 
The constraint is that the burden to 
the patient, $ep$ should not exceed $0.8$. The minimum is $0.665$ (which fortunately implies that 
with this treatment schedule the infection can be cleared successfully) and it is achieved at $(e,p)=(1,0.8)$. 
In other words, the drug should be $100\%$ efficient while it is active. 
This is illustrated in Figure $\ref{num1}$, which depicts the spectral radius $\rho \left(\Phi(e,0.8/e) \right)$.

\begin{figure}
\centerline{
\includegraphics[width=7cm]{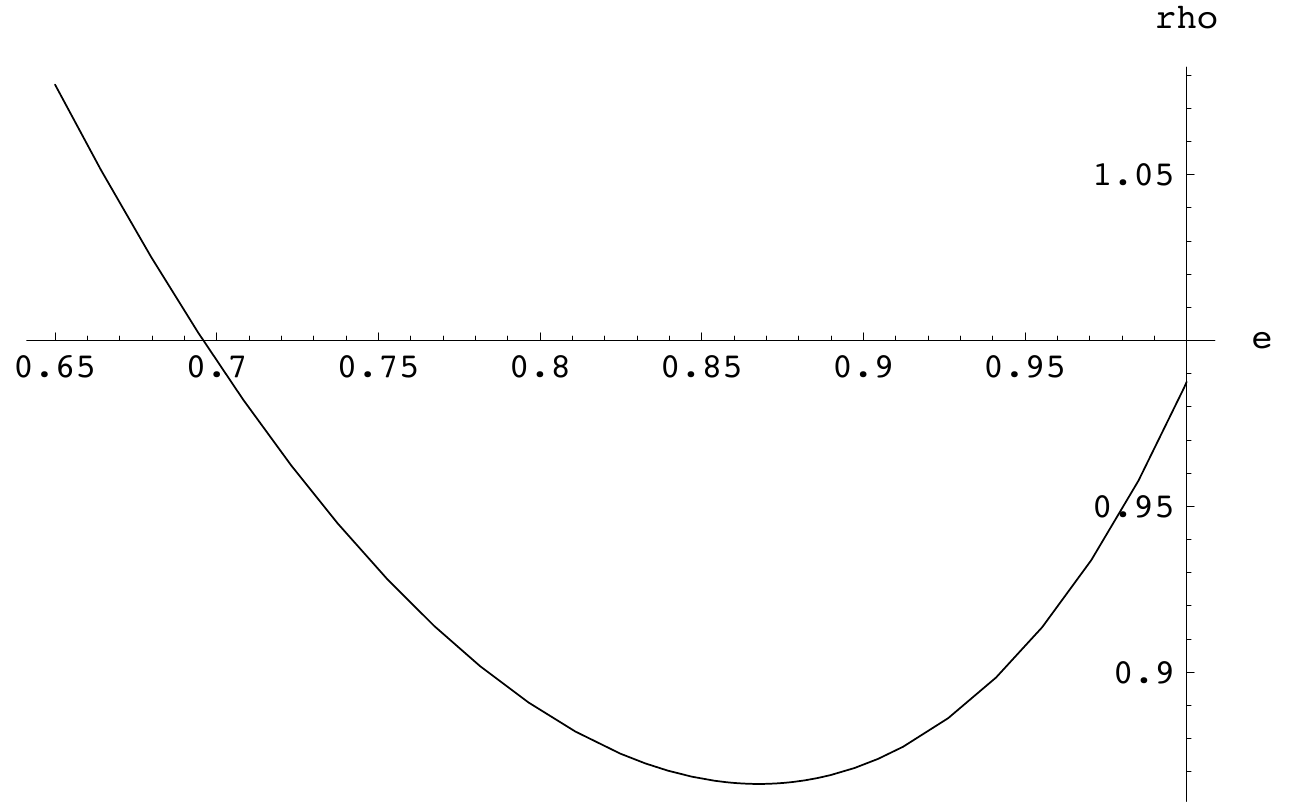}}
\caption{Graph of $\rho \left( \Phi(e,p)\right)$ for points $(e,p)$ with $e^2+p^2=1.2^2$ 
for $e\in[0.65,1]$. Minimum for $\rho$ is $0.866$ and it is achieved for $(e,p)=(0.868,0.829)$.}\label{num2}
\end{figure}

Let us see what happens when we modify the measure of the patient's burden to $e^2+p^2$, and demand that 
it should not exceed $1.2^2$. This time the minimal spectral radius 
is $0.866$ (again implying that this therapy will clear 
the infection) and it is achieved at $(e,p)=(0.868,0.829)$. 
This is illustrated in Figure $\ref{num2}$, which depicts the spectral radius 
$\rho \left(\Phi(e,\sqrt{1.2^2-e^2}) \right)$. 
A striking difference between this schedule 
and the previous one, is that now the minimum is achieved in the interior of the rectangular parameter space 
$[0,1]\times [0,1]$, while previously it was achieved on the boundary. 
When the drug is active, it should therefore not be $100\%$ efficient as before.

\begin{figure}
\centerline{
\includegraphics[width=8cm]{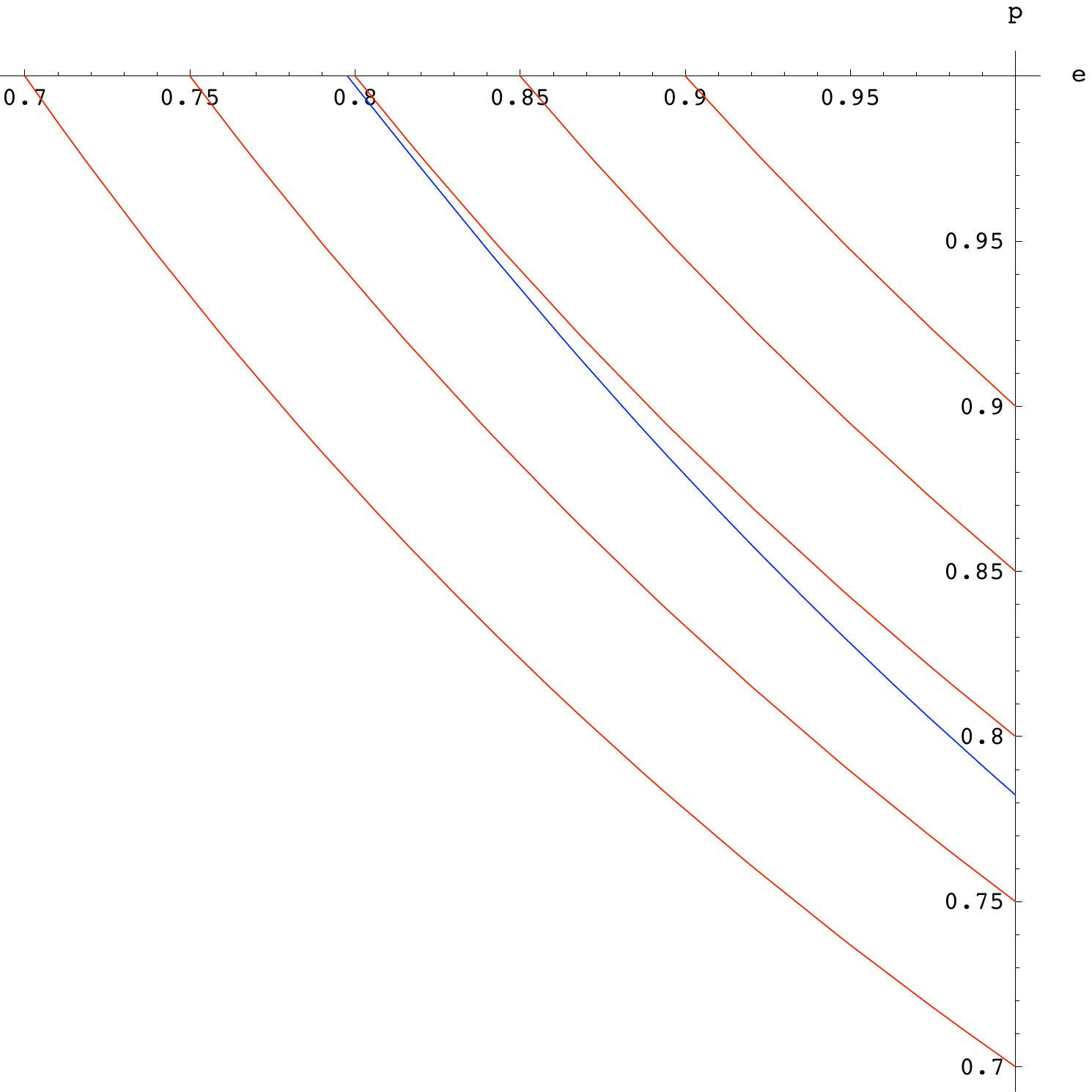}}
\caption{Level curves $ep=c$ in red ($c=0.7,0.75,0.8,0.85,0.9$, $c$ increasing in NE direction). 
Level curve $\rho \left(\Phi(e,p) \right)=0.7$ in blue.}\label{num3}
\end{figure}

Let us now consider minimization problems in which the patient's burden is minimized subject to 
a constraint on the spectral radius, or equivalently, on the speed of convergence to the 
infection-free equilibrium. If the patient's burden is measured by $ep$, and if the spectral radius 
should not exceed $0.7$, we find that the minimum is $0.782$ and it occurs at $(e,p)=(1,0.782)$ which 
is on the boundary of $[0,1]\times [0,1]$ and requires that the drug is $100\%$ effective when it is active.
This is illustrated in Figure $\ref{num3}$, where we depict a few level curves of $ep$, and the 
maximally allowable spectral radius $\rho=0.7$.

\begin{figure}
\centerline{
\includegraphics[width=8cm]{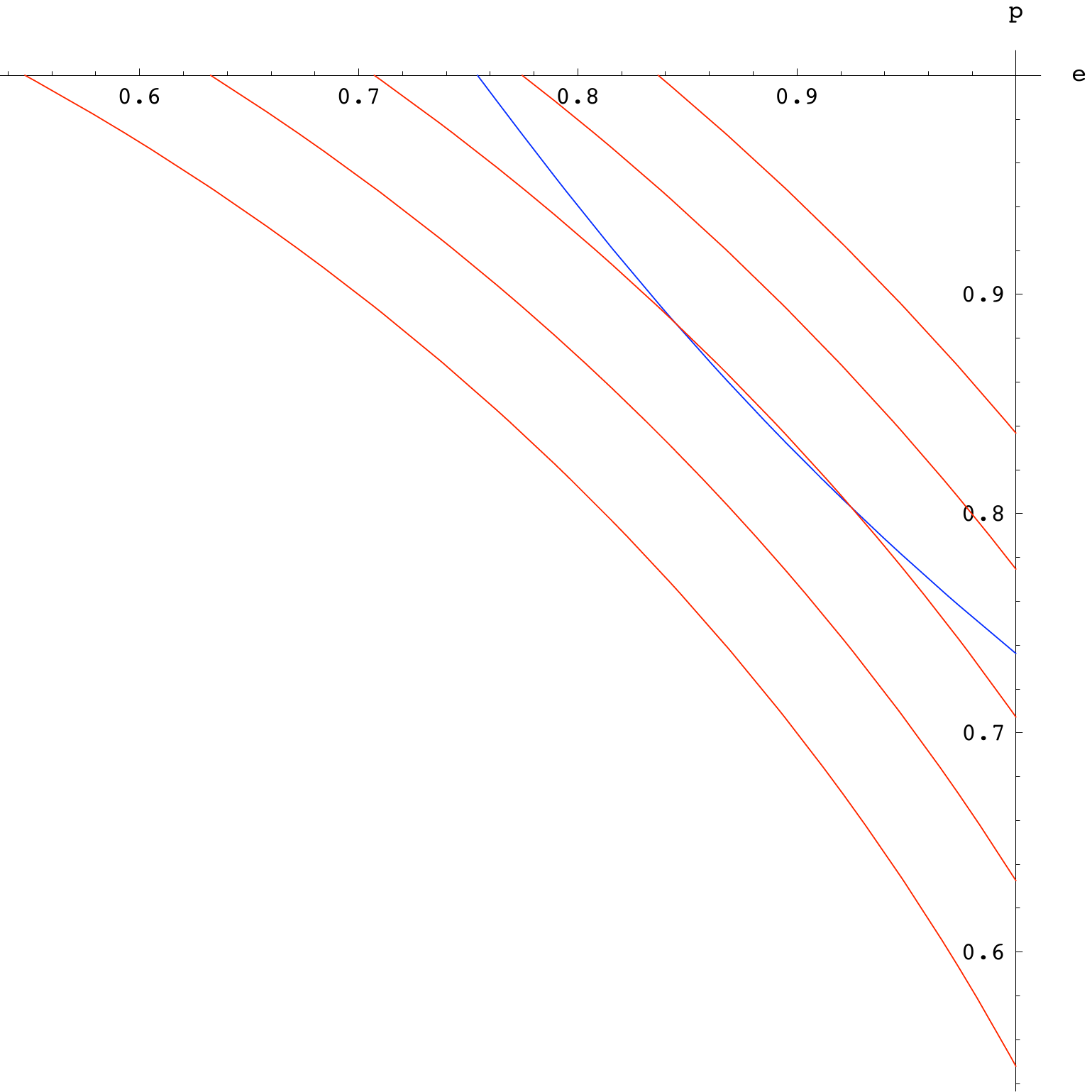}}
\caption{Level curves $e^2+p^2=c$ in red ($c=1.4,1.5,1.6,1.7,1.8$, $c$ increasing in NE direction). 
Level curve $\rho \left(\Phi(e,p) \right)=0.7$ in blue.}\label{num4}
\end{figure}

If we modify the measure of the burden to $e^2+p^2$ (and still assuming the constraint that the spectral 
radius should not exceed $0.7$), then the minimum is $1.493$ and it occurs at $(e,p)=(0.884,0.844)$ which 
is in the interior of the rectangular region $[0,1]\times[0,1]$. This is illustrated in Figure $\ref{num4}$, 
where we depict a few level curves $e^2+p^2$, and the maximally allowable spectral radius $\rho=0.7$.

\section*{Appendix}
\subsection*{Proof of Proposition $\ref{mono}$}
This proof hinges on the following standard facts:
\begin{enumerate}
\item If $A$ is quasi-positive and $A\leq B$ but $A\neq B$, then $0\leq \exponent^{tA}\leq \exponent^{tB}$ but $\exponent^{tA}\neq \exponent^{tB}$ for all $t>0$.

To see this, let $\alpha>0$ be such that $C=A+\alpha I\geq 0$. Then setting $D=B+\alpha I$, we have that 
$0\leq C\leq D$ but $C\neq D$. Then $\exponent^{tC}\leq \exponent^{tD}$, but $\exponent^{tC}\neq 
\exponent^{tD}$ for all $t>0$. It follows that $\exponent^{tA}\leq \exponent^{tB}$ but 
$\exponent^{tA}\neq \exponent^{tB}$ for all $t>0$. 

\item For all $t>0$, $\exponent^{tA(0)}>0$ if $e\neq 1$, 
while $\exponent^{tA(e)}\geq 0$ (but not $>0$) if $e=1$.

\item If $A>0$ and $B\geq 0$ has no zero row or zero column, then $AB>0$ and $BA>0$. 
This is true in particular when $B=\exponent^{tC}$ for $t\geq 0$ and $C$ a quasi-positive matrix because of Fact 1 and the fact that matrix exponentials are invertible. 

\item If  $0<A\leq B$ but $B\neq A$, then $\rho(A)<\rho(B)$, see Corollary $1.5$ in Chapter $2$ in \cite{berman-plemmons}.

\end{enumerate}
Continuity of the map $(e,p)\rightarrow \rho\left(\Phi(e,p)\right)$ 
follows from the definition $(\ref{def-fi})$ of 
$\rho$ and the fact that the spectral radius of any matrix is continuous in terms of its entries.

Let $0\leq e<e'<1$ and $p\neq 0$. Then:
\begin{eqnarray*}
&&A(e')\leq A(e)\textrm{ and }A(e')\neq A(e)\\
&\Rightarrow&0< \exponent^{pA(e')}\leq \exponent^{pA(e)}
\textrm{ and }\exponent^{pA(e')}\neq \exponent^{pA(e)}\textrm{ by Facts 1 and 2}\\
&\Rightarrow&0< \exponent^{(\tau-p)A(0)}\exponent^{pA(e')}\leq 
\exponent^{(\tau-p)A(0)}\exponent^{pA(e)}
\textrm{ and }\exponent^{(\tau-p)A(0)}\exponent^{pA(e')}\neq 
\exponent^{(\tau-p)A(0)}\exponent^{pA(e)}\textrm{ by Facts 1 and 3 }\\
&&\textrm{ and invertibility of matrix exponentials}\\
&\Rightarrow&0<  \Phi(e',p)\leq \Phi (e,p)\textrm{ and }
\Phi(e',p)\neq \Phi (e,p)\\
&\Rightarrow& \rho \left(\Phi(e',p)\right)< \rho \left(\Phi (e,p)\right)\textrm{ by Fact 4}.
\end{eqnarray*}
This result remains valid if $e'=1$ because $\rho\left(\Phi(e,p)\right)$ is continuous. 
This establishes $(\ref{mon-eps})$.

Let $0\leq p<p'<\tau$ and $e\neq 0$. Then: 
\begin{eqnarray*}
&&A(e)\leq A(0)\textrm{ and }A(e)\neq A(0)\\
&\Rightarrow&0\leq  \exponent^{(p'-p)A(e)}\leq \exponent^{(p'-p)A(0)}\textrm{ and }
\exponent^{(p'-p)A(e)}\neq \exponent^{(p'-p)A(0)}\textrm{ by Fact 1}\\
&\Rightarrow&0\leq  \exponent^{(p'-p)A(e)}\exponent^{pA(e)}\leq 
\exponent^{(p'-p)A(0)}\exponent^{pA(e)}\textrm{ and }\\
&&\exponent^{(p'-p)A(e)}\exponent^{pA(e)}\neq \exponent^{(p'-p)A(0)}\exponent^{pA(e)}\textrm{ by Fact 1 and invertibility of exponentials}\\
&\Rightarrow&0<\exponent^{(\tau-p')A(0)}\exponent^{(p'-p)A(e)}\exponent^{pA(e)}\leq \exponent^{(\tau-p')A(0)}\exponent^{(p'-p)A(0)}\exponent^{pA(e)}\textrm{ and }\\
&&\exponent^{(\tau-p')A(0)}\exponent^{(p'-p)A(e)}\exponent^{pA(e)}\neq \exponent^{(\tau-p')A(0)}\exponent^{(p'-p)A(0)}\exponent^{pA(e)} 
\textrm{ by Fact 2 and}\\
&& \textrm{invertibility of exponentials}\\
&\Rightarrow&0<  \Phi(e,p')\leq \Phi (e,p)\textrm{ and }
\Phi(e,p')\neq \Phi (e,p)\\
&\Rightarrow& \rho \left(\Phi(e,p')\right)< \rho \left(\Phi (e,p)\right)\textrm{ by Fact 4}.
\end{eqnarray*}
This remains valid if $p'=\tau$ because $\rho\left(\Phi(e,p)\right)$ is continuous. 
This establishes $(\ref{mon-p})$.

Finally, it follows from our standing assumption $(\ref{standing})$, that the determinant of 
$A(0)$ is negative. Thus $A(0)$ has a positive eigenvalue which implies $(\ref{boundary})$. 
Also, $(\ref{boundary2})$ is immediate from $(\ref{A-mat})$.


\subsection*{Proof of Proposition $\ref{extension}$}
The same facts as in the proof of Proposition $\ref{mono}$ will be used.

Continuity of the map $(E,P)\rightarrow \rho\left(\Phi(E,P)\right)$ 
follows from the definition $(\ref{def-fi})$ of 
$\rho$ and the fact that the spectral radius of any matrix is continuous in terms of its entries.

Fix $i\in\{1,\dots,m+1\}$ and let $e_i<e_i'<1$. Then:
\begin{eqnarray*}
&&A(e_i')\leq A(e_i)\textrm{ and }A(e_i')\neq A(e_i)\\
&\Rightarrow&0< \exponent^{(p_i-p_{i-1})A(e'_i)}\leq \exponent^{(p_i-p_{i-1})A(e_i)}
\textrm{ and }\exponent^{(p_i-p_{i-1})A(e_i')}\neq \exponent^{(p_i-p_{i-1})A(e_i)}\textrm{ by Facts 1 and 2}\\
&\Rightarrow&0<  \Phi \left( {\tilde E}_i,P\right)\leq \Phi \left(E,P\right)\textrm{ and }
\Phi\left({\tilde E}_i,P\right)\neq \Phi \left(E,P\right)\textrm{ by Facts 1 and 3 }\\
&&\textrm{ and invertibility of matrix exponentials}\\
&\Rightarrow& \rho \left(\Phi\left({\tilde E}_i,P\right)\right)< \rho \left(\Phi \left(E,P\right)\right)\textrm{ by Fact 4}.
\end{eqnarray*}
This result remains valid if $i=1$ and $e_1'=1$ because $\rho\left(\Phi(E,P)\right)$ is continuous.
This establishes $(\ref{ext-eps})$.

Fix $j\in\{1,\dots,m\}$ and let $0<p_j<p_j'<\tau$. Since $e_{j+1}<e_j$, we have that 
\begin{eqnarray*}
&&A(e_j)\leq A(e_{j+1})\textrm{ and }A(e_j)\neq A(e_{j+1})\\
&\Rightarrow&0\leq  \exponent^{(p_j'-p_j)A(e_j)}\leq \exponent^{(p'_j-p_j)A(e_{j+1})}\textrm{ and }
\exponent^{(p_j'-p_j)A(e_j)}\neq \exponent^{(p_j'-p_j)A(e_{j+1})}\textrm{ by Fact 1}\\
&\Rightarrow&0\leq  \exponent^{(p_j'-p_j)A(e_j)}\exponent^{(p_j-p_{j-1})A(e_j)}\leq 
\exponent^{(p_j'-p_j)A(e_{j+1})}\exponent^{(p_j-p_{j-1})A(e_j)}\textrm{ and }\\
&&\exponent^{(p_j'-p_j)A(e_j)}\exponent^{(p_j-p_{j-1})A(e_j)}\neq 
\exponent^{(p_j'-p_j)A(e_{j+1})}\exponent^{(p_j-p_{j-1})A(e_j)}\textrm{ by Fact 1 and invertibility of exponentials}\\
&\Rightarrow&0<\exponent^{(p_{j+1}-p_j')A(e_{j+1})}\exponent^{(p_j'-p_{j-1})A(e_j)}\leq 
\exponent^{(p_{j+1}-p_j)A(e_{j+1})}\exponent^{(p_j-p_{j-1})A(e_j)}\textrm{ and }\\
&&\exponent^{(p_{j+1}-p_j')A(e_{j+1})}\exponent^{(p_j'-p_{j-1})A(e_j)}\neq 
\exponent^{(p_{j+1}-p_j)A(e_{j+1})}\exponent^{(p_j-p_{j-1})A(e_j)}\textrm{ by Facts 1 and 3 and }\\
&& \textrm{invertibility of exponentials}\\
&\Rightarrow&0<  \Phi(E,{\tilde P}_j)\leq \Phi (E,P)\textrm{ and }
\Phi(E,{\tilde P}_j)\neq \Phi (E,P)\textrm{ by Facts 1 and 3 and invertibility of exponentials}\\
&\Rightarrow& \rho \left(\Phi(E,{\tilde P}_j)\right)< \rho \left(\Phi (E,P)\right)\textrm{ by Fact 4}.
\end{eqnarray*}
This establishes $(\ref{ext-p})$.


\begin{thebibliography}{199}
\bibitem{seema} S.H. Bajaria, G. Webb, and D.E. Kirschner, Predicting differential responses to structured treatment interruptions during HAART, Bulletin of Mathematical Biology 66, 1093-1118, 2004.

\bibitem{ball} C.L. Ball, M.A. Gilchrist, and D. Coombs, Modeling within-host evolution of HIV: mutation, competition and strain replacement, Bulletin of Mathematical Biology 69, 2361-2385, 2007.

\bibitem{berman-plemmons} A. Berman, and R. Plemmons, Nonnegative matrices in the 
mathematical sciences, SIAM, 1994. 

\bibitem{bonhoeffer97} S. Bonhoeffer, and M.A. Nowak, Pre-existence and emergence of drug resistance in HIV-1 infection, Proceedings of the Royal Society of London B 264, 631-637, 1997.

\bibitem{MathMed} P. De Leenheer, and S.S. Pilyugin, Multi-strain virus dynamics with mutations: a global analysis, to appear in Mathematical Medicine and Biology (Preliminary version in arXiv:0707.4501/).

\bibitem{SIAP03} P. De Leenheer, and H.L. Smith, Virus dynamics: a global analysis, SIAM Journal on Applied Mathematics 63, 1313-1327, 2003.

\bibitem{dixit} N. M. Dixit, and A.S. Perelson, Complex patterns of viral load decay under antiretroviral therapy: influence of pharmacokinetics and intracellular delay, Journal of Theoretical Biology 226, 95-109 (2004).

\bibitem{donofrio} A. d'Onofrio, Periodically varying antiviral therapies: conditions for global stability of the virus free state, Applied Mathematics and Computation 168, 945-953, 2005.



\bibitem{kirschner} D. Kirschner, S. Lenhart, and S. Serbin, Optimal control of the chemotherapy of HIV, 
Journal of Mathematical Biology 35, 775-792, 1997.


\bibitem{krakovska} O. Krakovska, and L.M. Wahl, Drug-Sparing Regimens for HIV Combination Therapy: Benefits predicted for "drug coasting", Bulletin of Mathematical Biology 69, 2627-2647, 2007.

\bibitem{nowak} M.A. Nowak, and R.M. May, Virus Dynamics, Oxford University Press, New York, 2000.

\bibitem{ortiz} G.M. Ortiz etal, Structured antiretroviral treatment interruptions in chronically HIV-1-infected subjects, Proceedings of the National Academy of Sciences 98, 13288-13293, 2001.

\bibitem{perelson} A.S. Nelson, and P.W. Nelson, Mathematical analysis of HIV-1 dynamics in vivo, SIAM Review 41, 3Ð44, 1999.

\bibitem{ribiero00} R.M. Ribiero, and S. Bonhoeffer, 
Production of resistant HIV mutants during antiretroviral therapy, Proceedings of the National 
Academy of Sciences 97, 7681-7686, 2000.

\bibitem{rong} L. Rong, Z. Feng, and A.S. Perelson, Drug resistance during antiretroviral treatment, 
Bulletin of Mathematical Biology 69, 2027-2060, 2007.

\end{thebibliography}
\end{document}